\documentclass[12pt]{article}

\usepackage{moreverb,url,array,tabularx,multirow,tablefootnote,graphicx,algorithm,algpseudocode,arydshln,tablefootnote,setspace,geometry,booktabs,amsmath,threeparttable,microtype}
\usepackage[symbol]{footmisc}
\usepackage[colorlinks,bookmarksopen,bookmarksnumbered,citecolor=red,urlcolor=red]{hyperref}

\newcommand\BibTeX{{\rmfamily B\kern-.05em \textsc{i\kern-.025em b}\kern-.08em
T\kern-.1667em\lower.7ex\hbox{E}\kern-.125emX}}

\usepackage{amsmath}
\usepackage{amsfonts}
\usepackage{amssymb}
\usepackage{graphicx}
\usepackage{verbatim}
\usepackage[super,comma]{natbib}
\immediate\write18{texcount -tex -sum  \jobname.tex > \jobname.wordcount.tex}

\author{
  Julia Gallini,
  Zach Baucom,
  Yorghos Tripodis
}
\title{Cognitive Factor-Based Selection Increases Power in Alzheimer’s Dementia Randomized Clinical Trials}

\begin{document}
\maketitle

\begin{abstract}
Alzheimer's dementia (AD) is of increasing concern as populations achieve longer lifespans. Many of the recent failed AD clinical trials had a low number of AD events and were thus underpowered. Previous trials have attempted to address this issue by requiring signs of cognitive decline in brain imaging for trial enrollment. However, this method systematically excludes people of color and those without access to healthcare and results in a selected sample that is not representative of the target patient population. We therefore propose the use of a predictive model based on cognitive test scores to enroll cognitively normal yet high risk participants in a hypothetical clinical trial. Cognitive test scores are a widely accessible tool so their use in enrollment would be less likely to exclude marginalized populations than biomarkers, such as imaging, which are overwhelmingly available to exclusively high-income patients. We developed a novel longitudinal factor model to predict AD conversion within a 3-year window based on data from the National Alzheimer's Coordinating Center. Through simulation we demonstrate that our predictive model provides substantial improvements in statistical power and required sample size in hypothetical clinical trials across a range of drug effects when compared to other methods of subject selection.
\end{abstract} \hspace{10pt}

\clearpage
\pagenumbering{arabic}

\section{Introduction}
Alzheimer's dementia (AD) is the most prevalent type of dementia and is becoming progressively more common as world populations achieve longer lifespans \cite{refadreport}. While the growing need for treatment options has been evident in scientific communities for decades, as of mid-2024 there are only two FDA approved drugs to slow progression of mild AD and four drugs to treat symptoms of AD (without slowing progression) \cite{mementine,lecanemab,donanemab,galantamine,rivastigmine,donepazil}. For each of these drugs there are a variety of significant concerns related to side effects, limited efficacy, and high cost \cite{earlyconcerns,pharmaceutics14061117,lecconcerns}. There thus remains a substantial gap between the supply and demand for  safe, affordable, and efficacious AD treatments.

There are a variety of reasons for the failure of the recent drug trials for AD \cite{ASHER2022120861}. One such reason is a lack of statistical power resulting from a low number of events (e.g. conversion to AD) during the trial. To attempt to address this concern, previous clinical trials have used strict inclusion and exclusion criteria based on combinations of demographic information, baseline cognitive tests, biomarkers, and imaging data \cite{giacobini2013alzheimer,selkoe2016amyloid}. However, clinical trials have yet to employ predictive modeling as a method for enrolling high-risk patients, a highly effective option for improving statistical power as demonstrated by \citet{ezzati2020machine}. Additionally, while imaging and biomarker data can be highly predictive of risk of AD, these methods are less likely to be accessible to minoritized groups \cite{Ricard}. Requiring such data to be available for an individual to enroll in a clinical trial results in relatively wealthy and predominantly white trial participants, limiting the generalizability of results to other populations. This bias is a particular problem given the evidence of already existing racial and socioeconomic differences in prevalence, pathology, and presentation of AD \citep{BABULAL2019292,GRAFFRADFORD2016669}. Specifically, multiple studies have found evidence of differential exclusion of racial minority groups from AD trials based on amyloid biomarker eligibility \citep{molina2024racial,raman2021disparities,grill2024eligibility}.  Further, many trials require that participants already be in a state of mild cognitive impairment (MCI) to enroll to improve power \citep{COX2019322}, when ideally AD drugs would be tested on a cognitively normal population and prevent the onset of any symptoms of cognitive impairment. 

Cognitive test scores are more accessible predictors of AD risk than imaging and biomarker data. They are simple and inexpensive to administer and are thus available to a more diverse population. We therefore propose a novel statistical method of extracting statistical factors from cognitive test scores to be used in a predictive model of three-year AD risk in cognitively normal patients. We then demonstrate through simulation the improved statistical power and required sample size using our cognitive factor model compared to other subject selection methods in a hypothetical AD randomized clinical trial.

\section{Methods}
\subsection{Data}
\subsubsection{Data Source}
The data for this project were downloaded from the National Alzheimer's Coordinating Center (NACC) Uniform Data Set (UDS) on 12/11/2023. NACC is a publicly available, National Institute on Aging funded, centralized repository of harmonized data from approximately 30 Alzheimer's Disease Research Centers (ADRC) in the United States \citep{NACC}. The overarching goal of the NACC is to facilitate research on AD and related diseases. Each ADRC follows a cohort of participants with and without cognitive impairment and approximately annually conducts harmonized cognitive, neuropsychiatric, and neurological evaluations. The ADRCs then deposit the harmonized data to the NACC which is compiled into large amounts of serial neuropsychological measurements. Analyzing these longitudinal outcomes provides vital insight into the disease course and presentation.

\subsubsection{Cohort Development}
Figure \ref{dataflow} displays a flow chart of the data cleaning process. The main variables of interest were 10 cognitive tests that are standard to administer for dementia evaluation (Immediate Recall, Delayed Recall, Digit Span Forward, Digit Span Backward, Animal List Generation, Vegetable List Generation, Boston Naming Test, Trail Making Test Part A, Trail Making Test Part B, Digit Symbol) \citep{bilgel2017temporal} and the covariates of interest (sex, education, age at baseline, race, APOE4, hypertension, diabetes, smoking years, binary BMI (obese/non-obese), TBI history (present/absent), and depression (ever/never)). Participants with any missing values for any of these variables were removed. Participants with data at only one time point were removed. Participants had to be classified as cognitively normal for at least two of their first visits to be included. 

The 10 cognitive test variables were standardized so that effect sizes could be compared. Since digit span forward and backward were the only two tests in which a higher score is a worse cognitive outcome, these values were multiplied by -1 prior to standardization so that their values were interpreted the same way as the other eight tests.

\subsection{Longitudinal Factor Model}
Factor scores allow for dimensionality reduction when dealing with large numbers of potential predictors like cognitive tests. We propose the following longitudinal factor model:

\begin{equation*}
\begin{aligned}
\boldsymbol{y_{ij}} = 
\begin{bmatrix}
y_{ij1}\\
y_{ij2}\\
\vdots\\
y_{ijK}
\end{bmatrix}
&= G \begin{bmatrix}
\alpha_{ij1}\\
\alpha_{ij2}\\
\vdots\\
\alpha_{ijQ}
\end{bmatrix}
+ 
\begin{bmatrix}
\varepsilon_{ij1}\\
\varepsilon_{ij2}\\
\vdots\\
\varepsilon_{ijK}
\end{bmatrix},  
\begin{bmatrix}
\varepsilon_{ij1}\\
\varepsilon_{ij2}\\
\vdots\\
\varepsilon_{ijK}
\end{bmatrix} 
\sim N(0, \Sigma_\varepsilon
)\\
\boldsymbol{\alpha_{ij}} = 
\begin{bmatrix}
\alpha_{ij1}\\
\alpha_{ij2}\\
\vdots\\
\alpha_{ijQ}
\end{bmatrix} & = 
\begin{bmatrix}
\alpha_{i(j-1)1}\\
\alpha_{i(j-1)2}\\
\vdots\\
\alpha_{i(j-1)Q}
\end{bmatrix} +
\begin{bmatrix}
\eta_{ij1}\\
\eta_{ij2}\\
\vdots\\
\eta_{ijQ}
\end{bmatrix}, 
\begin{bmatrix}
\eta_{ij1}\\
\eta_{ij2}\\
\vdots\\
\eta_{ijQ}
\end{bmatrix} \sim N(0, \delta_{ij}\Sigma_\eta)
\end{aligned}
\end{equation*}

\noindent where \(y_{ijk}\) is the \(k^{th}\) neuropsychological test (\(k\in\{1, 2, ..., K\}\)) at the \(j^{th}\) observation (\(j \in \{1, 2, ..., J\}\)) for the \(i^{th}\) subject (\(i \in \{1, 2, ..., N\}\)). Without loss of generality, this model assumes that there are no baseline covariate effects \(x_{ij}\beta\) in the model. The model can be easily extended to include these effects by first estimating the linear effects \(x_{ij}\beta\) on \(y_{ij}\), subtracting these estimates from the original \(y_{ij}\), and proceeding to estimate the factor model above without the \(x_{ij}\beta\) component and using the updated, linearly adjusted \(y_{ij}^*\) values \citep{harvey1990forecasting}. In this paper we fit the extended version of this model using the covariates described in section 2.1.2. We also include a factor loading matrix \(\boldsymbol{G} \in R^{K\times Q}\). The matrix \(\boldsymbol{G}\) transforms the underlying state vector \(\boldsymbol{\alpha_{ij}}\) from a \(Q\times 1\) vector back into a \(K\times 1\) vector, with $(Q<K$). The \(\boldsymbol{\alpha_{ij}}\) vector can be thought of as latent cognitive factors that we will use later in the paper to predict AD outcomes. The \(\Sigma_\epsilon\) matrix estimates the measurement error variance structure. The \(\Sigma_\eta\) matrix estimates the variance structure of the states themselves, while the scalar \(\delta_{ij}\) reflects the difference in unit time between observations \(j-1\) and \(j\) and adjusts the variance matrix appropriately \citep{durbin_koopman}.

This model can be estimated within a Bayesian framework using a Gibbs sampler.  The steps of the Gibbs sampler are:
\begin{enumerate}
  \item Estimate \(\boldsymbol{\alpha_{ij}}\) using the Kalman filter and smoother conditioned on priors for \(\boldsymbol{G},\Sigma_\epsilon\), and \(\Sigma_\eta\).
  \item Estimate \(\boldsymbol{G}\) from its posterior conditioned on the estimate for \(\boldsymbol{\alpha_{ij}}\) and priors for \(\Sigma_\epsilon\) and \(\Sigma_\eta\).
  \item Estimate \(\Sigma_\epsilon\) from its posterior conditioned on the estimates for \(\boldsymbol{\alpha_{ij}}, \boldsymbol{G}\), and the prior for \(\Sigma_\eta\).
  \item Estimate \(\Sigma_\eta\) from its posterior conditioned on the estimates for \(\boldsymbol{\alpha_{ij}}, \boldsymbol{G}\), and \(\Sigma_\epsilon\).
  \item Repeat steps 1-4 using the most recent estimates of the unknown parameters each time until the estimates for each parameter converge.
\end{enumerate}

In the following sections, we describe in detail the derivation of the conditional posterior distributions for each parameter.

\subsubsection{Estimation of the \(\alpha_{ij}\) vector}
Estimating \(\boldsymbol{\alpha_{ij}}\) can be achieved through a forward Kalman filter and backward Kalman smoother sampler \citep{durbin_koopman}. The Kalman filter and smoother recursions can be computed independently for each subject \(i \in \{1, 2, ..., N\}\). 

Let $\psi=\{\boldsymbol{G},\Sigma_\epsilon,\Sigma_\eta\}$ denote the vector of the other unknown parameters. We select a prior for \(\boldsymbol{\alpha_{ij}}\) for all \(i,j\): \(P(\boldsymbol{\alpha_{ij}})\sim N(\boldsymbol{m_0},\boldsymbol{P_0})\), where \(\boldsymbol{m_0}\) is the prior mean and \(\boldsymbol{P_0}\) is the prior variance.
\noindent Based on the prior defined for \(\boldsymbol{\alpha}_{ij}\), using the Kalman Filter we can calculate \(\boldsymbol{\alpha}_{ij}|\boldsymbol{y}_{i(1:j)} \sim N_\psi(\boldsymbol{\alpha}_{ij|ij}, \boldsymbol{P}_{ij|ij})\) and \(\boldsymbol{\alpha}_{i(j+1)}|\boldsymbol{\alpha}_{ij} \sim N_\psi(\boldsymbol{\alpha}_{ij}, \Sigma_\eta)\), where \(\boldsymbol{\alpha}_{ij|ij}\) is the updated state estimate, \(\boldsymbol{P}_{ij|ij}\) is the updated state estimate variance, \(\boldsymbol{\alpha}_{ij}\) is the mean of the distribution of the predicted estimate for \(\boldsymbol{\alpha}_{i(j+1)}|\boldsymbol{\alpha}_{ij}\), and \(\Sigma_\eta\) is the variance of this estimate. Because of normality, the posterior distribution for \(\boldsymbol{\alpha}_{ij}\) after running the forward Kalman filter is \(N(\boldsymbol{m}_{ij}, \boldsymbol{R}_{ij})\), where \(\boldsymbol{m}_{ij} = E_\psi(\boldsymbol{\alpha}_{ij}| \boldsymbol{\alpha}_{i(j+1)},\boldsymbol{y}_{i(1:j)})\) and \(\boldsymbol{R}_{ij} = \text{Var}_\psi(\boldsymbol{\alpha}_{ij}| \boldsymbol{\alpha}_{i(j+1)},\boldsymbol{y}_{i(1:j)})\) \citep{shumway_stoffer}.  

For the backward smoother sampling procedure we start by sampling \(\boldsymbol{\alpha}_{iJ^*}\) from a \(N_\psi(m_{iJ}, \boldsymbol{R}_{iJ})\) distribution. We then set \(\boldsymbol{\alpha}_{iJ^*} = \boldsymbol{\alpha}_{iJ}\) for the calculation of \(\boldsymbol{m}_{i(J-1)}\). Then we sample \(\boldsymbol{\alpha}_{i(J-1)}^*\) from a \(N_\psi(\boldsymbol{m}_{i(J-1)}, \boldsymbol{R}_{i(J-1)})\) distribution. This process continues until a whole chain \(\boldsymbol{\alpha}_{i(0:J)}^*\) has been sampled. We now have smoothed posterior estimates for \(\boldsymbol{\alpha}_{ij}\). More details on these algorithms can be found in \citet{shumway_stoffer} and \citet{durbin_koopman}.

\subsubsection{Estimation of \(G\) Matrix}
Conditioned on the \(\boldsymbol{\alpha_{ij}}\) estimates acquired in the previous step and priors for the other variables we estimate the \(\boldsymbol{G}\) matrix. We denote each row of G as \(\boldsymbol g_{k}\) (dimension \(Q\times 1\)) for \(k\in\{1, 2, ..., K\}\), and give \(\boldsymbol g_{k}\) a multivariate normal prior:
 
\begin{equation*}
\begin{aligned}
P(\boldsymbol g_{k}) &= \frac{1}{\sqrt{2\pi \sigma^2_{gk} }}e^\frac{-(\boldsymbol g_{k} - \boldsymbol \mu_{k})^{T}(\boldsymbol g_{k} - \boldsymbol \mu_{k})}{\sigma^2_{gk}}\\
\end{aligned}
\end{equation*}

\noindent where \(\mu_k\) is the prior mean of \(g_k\) and \(\sigma_{gk}^2\) is the prior variance. After further algebraic manipulation we have:
\begin{equation*}
    \begin{aligned}
-2log(P(\boldsymbol{g_k}))&\propto [-2 \boldsymbol \mu_{k}^T\boldsymbol g_{k}+\boldsymbol g_{k}^T\boldsymbol g_{k}]/\sigma^2_{gk}.  
    \end{aligned}
\end{equation*}
\noindent The data likelihood with respect to \(g_{k}\) can be written as:
\begin{equation*}
\begin{aligned}
-2logP(Y|\boldsymbol g_{k}, \boldsymbol{\alpha},\Sigma_{\epsilon},\Sigma_\eta) &\propto \sum_{i, j} (y_{ijk} - \boldsymbol \alpha_{ij}^T\boldsymbol g_{k})^2/\sigma^2_{\varepsilon k} \\
&\propto \sum_{i, j} (-2y_{ijk}\boldsymbol \alpha_{ij}^T\boldsymbol g_{k} + \boldsymbol g_{k}^T\boldsymbol \alpha_{ij}\boldsymbol \alpha_{ij}^T\boldsymbol g_{k})/\sigma^2_{\varepsilon k}\\
&\propto [-2(\sum_{i, j} y_{ijk}\boldsymbol \alpha_{ij}^T)\boldsymbol g_{k} + \boldsymbol g_{k}^T(\sum_{i, j} \boldsymbol \alpha_{ij}\boldsymbol \alpha_{ij}^T)\boldsymbol g_{k})]/\sigma^2_{\varepsilon k}\\
\end{aligned}
\end{equation*}

\noindent where \(\sigma_{\epsilon k}^2\) is the \(k^{th}\) diagonal element of the \(\Sigma_\epsilon\) matrix. By combining the two we can write the posterior distribution as:
\begin{equation*}
\begin{aligned}
-2log(P(\boldsymbol g_{k}|Y,\boldsymbol{\alpha},\Sigma_{\epsilon},\Sigma_\eta))\propto & [-2 \boldsymbol \mu_{k}^T\boldsymbol g_{k}+\boldsymbol g_{k}^T\boldsymbol g_{k}]/\sigma^2_{gk} + [-2(\sum_{i, j} y_{ijk}\boldsymbol \alpha_{ij}^T)\boldsymbol g_{k}+ \boldsymbol g_{k}^T(\sum_{i, j}\boldsymbol  \alpha_{ij}\boldsymbol \alpha_{ij}^T)\boldsymbol g_{k})]/\sigma^2_{\varepsilon k}\\
\propto & [-2 (\sigma^2_{\varepsilon k}\boldsymbol  \mu_{k}^T + \sigma^2_{gk}\sum_{i, j} y_{ijk}\boldsymbol \alpha_{ij}^T) \boldsymbol g_{k}+\boldsymbol g_{k}^T(\sigma^2_{gk} \boldsymbol (\sum_{i,j} \alpha_{ij}\boldsymbol \alpha_{ij}^T)+\sigma^2_{\varepsilon k} \boldsymbol I)\boldsymbol g_{k}]/\sigma^2_{gk}\sigma^2_{\varepsilon k}\\
\propto & (\boldsymbol{g}_k-\boldsymbol{R})^T\frac{\Sigma_\alpha}{\sigma^2_{gk} \sigma^2_{\varepsilon k}}(\boldsymbol{g}_k-\boldsymbol{R})\\
\end{aligned}
\end{equation*}

\noindent where \(\boldsymbol{R}=\boldsymbol \Sigma_\alpha^{-1} (\sigma^2_{\epsilon k}\boldsymbol \mu_{k}^T + \sigma^2_{gk} \sum_{i, j} y_{ijk}\boldsymbol \alpha_{ij}^T)\) and
\(\boldsymbol \Sigma_\alpha = (\sigma^2_{gk} \boldsymbol (\sum_{i,j}\boldsymbol\alpha_{ij}\boldsymbol\alpha_{ij}^T)+\sigma^2_{\varepsilon k} \boldsymbol I)\). The posterior distribution is thus denoted as \(\boldsymbol{g}_{k}|... \sim N(\boldsymbol R, \boldsymbol \Sigma^{-1}_\alpha\sigma^2_{gk}\sigma^2_{\varepsilon k})\).

Lastly, to estimate \(\boldsymbol G\) we must decide which tests load on to certain state processes. To make this decision we propose a structured approach, though an unstructured approach is also feasible with this model. For the structured approach, deciding which tests load onto which factors requires clinical subject matter expertise. Based on work by \citet{hayden2011factor} we decided on four statistical factors each representative of a different cognitive domain: memory, working memory, language and psychomotor speed ability. The 10 cognitive tests are chosen to load on the factor/cognitive domain they are meant to estimate. For this analysis test scores each load onto only one of the four proposed factors: immediate recall and delayed recall load onto the memory factor; digit span forward and digit span backward load onto the working memory factor; animal list generation, vegetable list generation, and Boston naming test load onto the language factor; trail making test part A, trail making test part B, and digit symbol load onto the psychomotor speed factor.

At the end of each iteration of the Gibbs sampler a random value for \(\boldsymbol{g_k}\) is drawn from the posterior distribution. Once all iterations have completed the element-wise mean of each \(\boldsymbol{g_k}\) is taken over all iterations (excluding burn-in) and combined to form the final estimate of \(\boldsymbol{G}\).

\subsubsection{Estimation of \(\Sigma_\epsilon\) and \(\Sigma_\eta\)}

Next in the Gibbs sampler we estimate the variance components conditioned on previous estimates of the other parameters. We now derive their posterior distributions. Fixing \(\Sigma_\epsilon\) to be a diagonal matrix, we estimate each of the \(k\) elements \(\sigma^2_{\epsilon k}\) independently: 
\begin{equation*}
    \begin{aligned}
P(\sigma^2_{\epsilon k})&\sim InvGamma(\frac{c_{0k}}{2},\frac{d_{0k}}{2})\\
P(\sigma^2_{\epsilon k}) &\propto {(\sigma_{\epsilon k}^2})^{-\frac{c_{0k}}{2}-1}e^{-\frac{d_{0k}}{2\sigma_{\epsilon k}^2}}\\
P(Y,\boldsymbol{G},\boldsymbol{\alpha},\Sigma_\eta|\sigma_{\epsilon k}^2) &\propto (\sigma_{\epsilon k}^2)^{\frac{-NJ}{2}}e^\frac{-\sum_{i,j}(y_{ij}-\boldsymbol{G\alpha}_{ij})^2}{2\sigma_{\epsilon k}^2}\\
P(\sigma_{\epsilon k}^2|Y,\boldsymbol{G},\boldsymbol{\alpha},\Sigma_\eta)&\propto P(y,\boldsymbol{G},\boldsymbol{\alpha}|\sigma_{\epsilon k}^2)P(\sigma_{\epsilon k}^2)\\
P(\sigma_{\epsilon k}^2|Y,\boldsymbol{G},\boldsymbol{\alpha},\Sigma_\eta)&\propto (\sigma_{\epsilon k}^2)^{\frac{-NJ+c_{0k}}{2}-1} e^\frac{-(d_{0k}+\sum_{i,j}(y_{ij}-\boldsymbol{G\alpha}_{ij})^2)}{2\sigma_{\epsilon k}^2}\\
\sigma_{\epsilon k}^2&\sim InvGamma\left(\frac{NJ+c_{0k}}{2},\frac{d_{0k}+\sum_{i,j}(y_{ij}-\boldsymbol{G\alpha}_{ij})^2}{2}\right).
    \end{aligned}
\end{equation*}

For \(\Sigma_\eta\), we fix the diagonal to be equal to 1 to standardize the components of the matrix and allow for comparison of factor scores across different studies. We allow for covariance between the \(\boldsymbol{\alpha_{ij}}\) values, which is also the correlation between factors because the components have been standardized. So unlike the estimation of \(\Sigma_\epsilon\) we estimate \(\Sigma_\eta\) using multivariate distributions:
\begin{equation*}
\begin{aligned}
\Sigma_\eta &\sim \text{Inv-Wishart}_{\nu_\eta}(\Lambda_\eta)\\
P(\Sigma_\eta) &\propto |\Sigma_\eta|^{-(\nu_\eta+p+1)/2}\text{exp}(-tr(\Lambda_\eta\Sigma_\eta^{-1}))\\
P(Y, \boldsymbol{G},\boldsymbol{\alpha}, \Sigma_\epsilon|\Sigma_\eta) &\propto |\Sigma_\eta|^{-(N(J-1))/2}\text{exp}(-\sum^N_{i = 1}\sum^{J}_{j= 2} (\boldsymbol{\alpha}_{ij} - \boldsymbol{\alpha}_{i(j-1)})'\Sigma_\eta^{-1}(\boldsymbol{\alpha}_{ij} - \boldsymbol{\alpha}_{i(j-1)})/2 \\
P(\Sigma_\eta|Y, \boldsymbol{G},\boldsymbol{\alpha}, \Sigma_\epsilon) & \propto P(Y, \boldsymbol{G},\boldsymbol{\alpha}, \Sigma_\epsilon|\Sigma_\eta)P(\Sigma_\eta) \\ 
&\propto |\Sigma_\eta|^{-(N(J-1) +\nu_\eta+p+1)/2}\text{exp}(-tr(\Lambda_\eta + \sum^N_{i = 1}\sum^J_{j= 2} (\boldsymbol{\alpha}_{ij} - \boldsymbol{\alpha}_{i(j-1)})^2)\Sigma_\eta^{-1})\\
\Sigma_\eta & \sim \text{Inv-Wishart}_{\nu_\eta + N(J-1)}(\Lambda_\eta+\sum^N_{i = 1}\sum^J_{j= 2} (\boldsymbol{\alpha}_{ij} - \boldsymbol{\alpha}_{i(j-1)})^2)
\end{aligned}
\end{equation*}
.

\noindent The covariance matrix \(\Sigma_\varepsilon\) and \(\Sigma_\eta\) both offer meaningful insight into underlying constructs of cognition. Additionally, these models allow for comparison of linear effects across different tests. Final estimates for these matrices are extracted in a similar fashion to previous parameters.

\subsubsection{Application of Model}
To ensure model convergence we ran 10,000 iterations of the Gibbs sampler.  The first 5,000 of these iterations were discarded as burn-in samples. 

For this analysis the main parameter of interest was the \(\boldsymbol{\alpha_{ij}}\) values (latent cognitive factor scores) and their predictive abilities of time to AD. We next describe an application of these latent cognitive factors relevant to clinical trials.

\subsection{Randomized Clinical Trial Simulation}

\begin{figure}[hbt!]
\centering
\includegraphics[scale=.92]{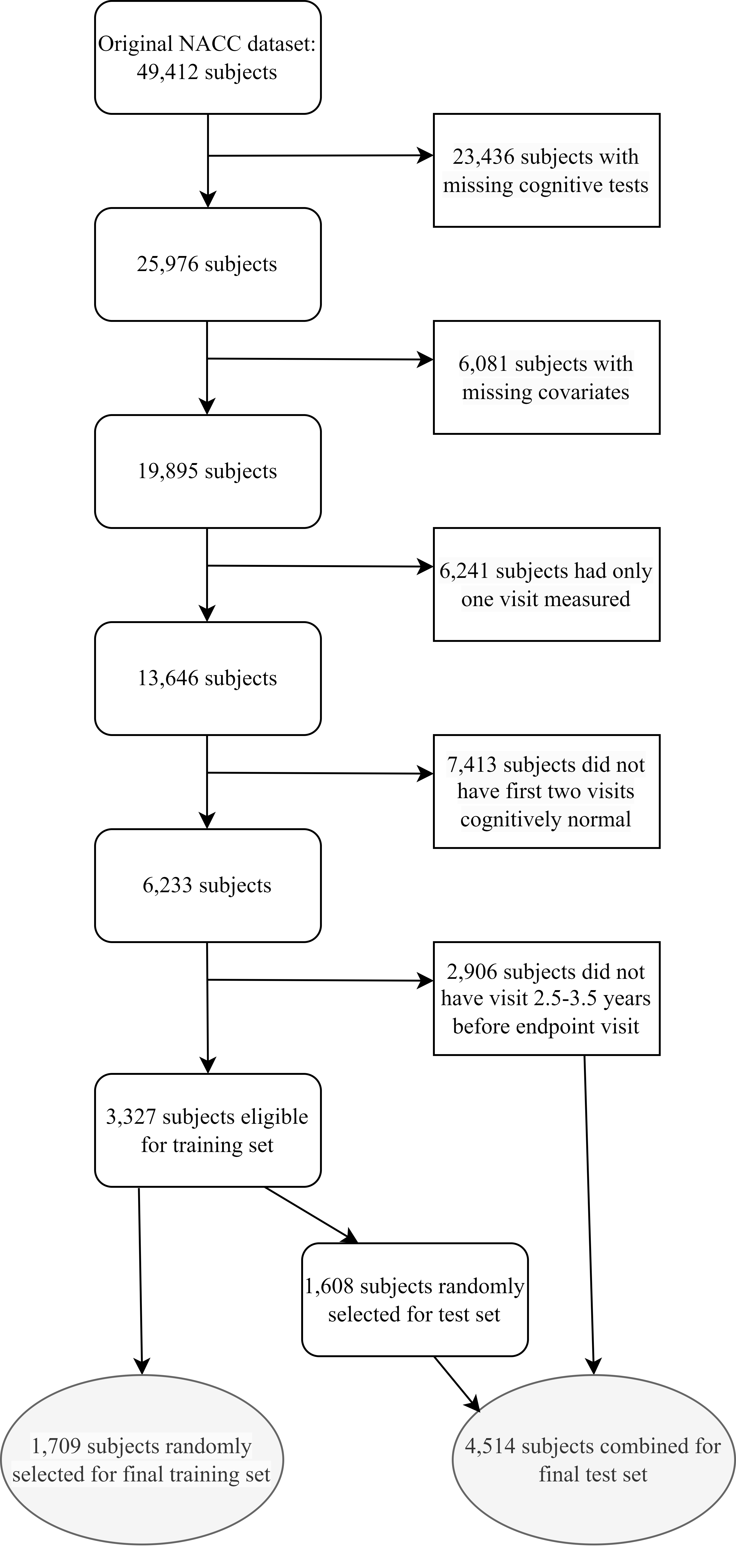}
\caption{Data Cleaning Flow Chart\label{dataflow}}
\end{figure}

To demonstrate the practical utility of latent cognitive factors we simulated a randomized clinical trial using a predictive cognitive factor model for subject selection from the NACC data. The participants were divided into training and test sets as described in Figure \ref{dataflow}. The training set was used for developing predictive models for AD and the test set was used for the actual RCT simulation including the power and required sample size calculations.

While the participants' ultimate AD status was used in the randomized clinical trial simulation, only the data from when the participants were cognitively normal was used for developing predictive models. The main outcome of interest for the predictive models was three-year ``objective decline" risk. We defined  ``objective decline" as either being classified as ``Dementia'' or ``Mild Cognitive Impairment'' using the cognitive status at UDS visit variable from the NACC dataset. To derive this variable we selected the first visit that a subject was classified as having MCI/Dementia or the last visit recorded for those who had not converted to MCI/Dementia as the ``endpoint visit''. We then looked backward in time for a visit with complete and cognitively normal data in the range of 2.5-3.5 years before the endpoint visit. If a participant had multiple visits in this range we selected the visit closest to 3 years from the endpoint visit. The data from this earlier visit was used in predictive modeling. Participants without a visit in this range were not considered for the training set to develop the predictive model. After cleaning, 3,327 participants were considered eligible for the training set. Each of these participants was randomly assigned to either the training or test set with probability 0.5 for each group. This resulted in 1,719 participants randomly selected for the predictive model training set, while the remaining 1,608 participants were combined to form the test set with the 2,906 participants that had at least two cognitively normal visits but did not have a visit in the three-year window before their endpoint visit (Figure \ref{dataflow}). These 2,906 participants were eligible for the test set because the simulated RCT looked forward in time from the second cognitively normal visit and did not require an endpoint visit in the three-year time window to allow for realistic lost to follow up and censoring. Summary statistics were calculated on demographic variables for test set participants.

The factor predictive model was developed using the training set. Four logistic regression models were created each using one of the four cognitive factors (memory, working memory, language, and psychomotor speed) and a standard set of covariates (sex, education, age at baseline, race, APOE4, hypertension, diabetes, smoking years, binary BMI, TBI history, and depression) as predictors and three-year AD status as the outcome. Ten more models were run for each of the ten original cognitive tests in place of the factors as predictors so that the predictive ability of the factors could be compared to the original tests. One logistic regression model with the same outcome and covariates was run including all four factors simultaneously to determine which of the factors were most statistically relevant. A final factor predictive model was selected from this model by removing cognitive factor predictors that were not statistically significant. All covariates remained in the final model regardless of significance.

Once the final factor model was selected we began subject selection for the simulation of a randomized clinical trial using the test set. Three different subject selection methods were explored for our simulation: random selection, factor model selection, and covariate-only model selection. Random selection was implemented by sampling 1,000 participants with replacement from the full test set of 4,514 participants. The factor model selection method was implemented by selecting 1,000 participants with replacement from a high-risk subset of the 4,514 test set participants. The high-risk subset was selected by predicting the probability of conversion to AD in three years using the predictive factor model. A threshold value of 0.15 provided the most balanced sensitivity and specificity for the factor model, so test set participants with probability of conversion greater than 0.15 were included in the high-risk subset (N=1,651) and other participants were excluded from consideration. The covariate-only model selection was performed similarly, except instead of using the factor model to predict probability of conversion to AD a model with only covariates and no cognitive factors was used. Participants with probability of conversion greater than 0.15 formed the high-risk set (N=1,659) and other participants were excluded; 1,000 participants were selected with replacement from this high-risk set. 1,248 participants appeared in both the factor model high-risk subset and the covariate model high-risk subset.

We used a time to MCI/Dementia framework to analyze the simulated RCT, specifically a Cox proportional hazards model. First, the truly observed (without the effect of a drug) time to MCI/Dementia outcomes were derived. The RCT baseline visit for each subject was the second observed cognitively normal visit. We looked forward in time $3.5$ years after baseline and excluded any visits after that range since a three year trial would be over at that point. For those that had an event and converted to MCI/Dementia within $3.5$ years we used their time from baseline to first MCI/Dementia visit as their follow up time in the Cox model. All other participants were censored either at their cognitively normal visit closest to three years from baseline or at their date of death if death occurred within the trial window and before MCI/Dementia conversion could occur. Participants who did not have a visit at all between baseline and 3.5 years later were lost to follow up.

We next simulated the effect of a drug on the true outcomes derived above. For each selection method participants were randomly assigned to the treatment or control arms with 0.5 probability of assignment to each arm. A range of treatment effects was implemented from 0.05 to 0.50. A new ``trial outcome'' was derived for each subject. Participants in the control arm maintained their true three year MCI/Dementia status and follow-up time as their trial outcome. Participants in the treatment arm that did not convert to AD also maintained their true three year MCI/Dementia status and follow-up time as their trial outcome, as it was assumed an MCI/Dementia drug would not be causing patients to develop MCI/Dementia who would not have otherwise. Participants in the treatment arm who converted to MCI/Dementia were assigned a (potentially) new trial outcome from a random binomial distribution with probability of MCI/Dementia conversion equal to 1 - treatment effect. Participants for whom the drug worked and thus were now censored were also assigned a new trial follow up time of exactly three years. Participants for whom the drug did not work maintained their original time to MCI/Dementia follow up time. Thus, the prevalence of MCI/Dementia events in the treatment arm was reduced by approximately the treatment effect, with sufficient randomness introduced to emulate the variability of the effects of a drug in a real-life trial.

A Cox proportional hazards model was run for each of the three selection options with treatment effect as main predictor. It was assumed most covariates would be relatively balanced due to randomization, but we adjusted for APOE4, education, and sex since these covariates are so highly associated with AD outcomes. 10,000 iterations of this simulation were performed for all 30 combinations of the three selection types and 10 treatment effects to ensure robust results. Iterations for which (by random chance) the hazard ratio for treatment effect was greater than 1 were given a power of 0 since such iterations would be unable to detect a protective effect of a treatment. These same iterations were also given an infinite required sample size to achieve 80\% power.

The median statistical power was calculated for each of the three models and each of the 10 treatment effects across all iterations. Since the standard sample size calculation for Cox models calculates the required number of events, to get the total required sample size for each iteration we took the required number of events for 80\% power and divided by the probability of an event (MCI/Dementia conversion) in the iteration. We calculated median required sample size for each of the three models and each of the 10 treatment effects across all iterations.

\section{Results}

\subsection{Prediction of Three-Year Alzheimer's Dementia Risk}

Demographics of the test set are displayed in Table \ref{tab1}. The test set was 84.1\% white and 34.5\% male. 28.3\% had ever had depression and 10.8\% had ever had a traumatic brain injury (TBI). 69.9\% had zero e4 alleles, and the average number of years of education was 15.8 (2.86). Demographic distributions in the test set were extremely similar to the training set (data not shown).

The results of the ten logistic regression models for three-year AD risk using each of the ten individual base cognitive tests as predictors are displayed in Table~\ref{tenmods}. The results of the four logistic regression models for three-year AD risk using each of the four factors as predictors are also displayed in Table~\ref{tenmods} directly below the respective tests that load onto each factor. All of these models were built solely on the training set. All models are adjusted for sex, education, age at baseline, race, APOE4, hypertension, diabetes, smoking years, binary BMI, TBI history, and depression. All of the cognitive tests were individually  statistically significant predictors of three-year AD risk except for Digit Span Backward; all of the factors were individually statistically significant. For the individual cognitive test models, model sensitivities ranged from 0.69 to 0.71 and specificities ranged from 0.65 to 0.70. Effect sizes ranged from 0.60 to 0.88 (for all tests, better scores were protective against AD). For the factor models, sensitivities ranged from 0.69 to 0.71 and specificities ranged from 0.67 to 0.68. Effect sizes ranged from 0.55 to 0.83. The observed improvements in prediction using factor scores instead of individual cognitive tests were quite small but present. Additional substantial advantages of the use of factor scores are described in the discussion. We thus proceeded with training our predictive model using factor scores instead of individual cognitive tests.

In a logistic regression model containing all four cognitive factors and all covariates as predictors of three-year AD risk only the memory and language factors were statistically significant at the 0.05 level. Therefore, our final predictive model used only these two factor scores as predictors in addition to the standard covariates (sensitivity=0.73, specificity=0.68). We used this model as our factor predictive model in the randomized controlled trial simulation; results from this simulation are described in the next section.

\subsection{Randomized Clinical Trial Simulation}

A set of 10,000 randomized controlled trials were simulated for theoretical treatment effects ranging from a 5\% reduction in risk of developing AD to a 50\% reduction. As we would expect to be the case in a real trial, the theoretical underlying treatment effect of a given drug is in many cases not the observed treatment effect in the trial. This phenomenon is due in part to the varied efficacy of drugs within individual participants. 

\begin{figure}[hbt!]
\centering
\includegraphics[width=0.85\textwidth]{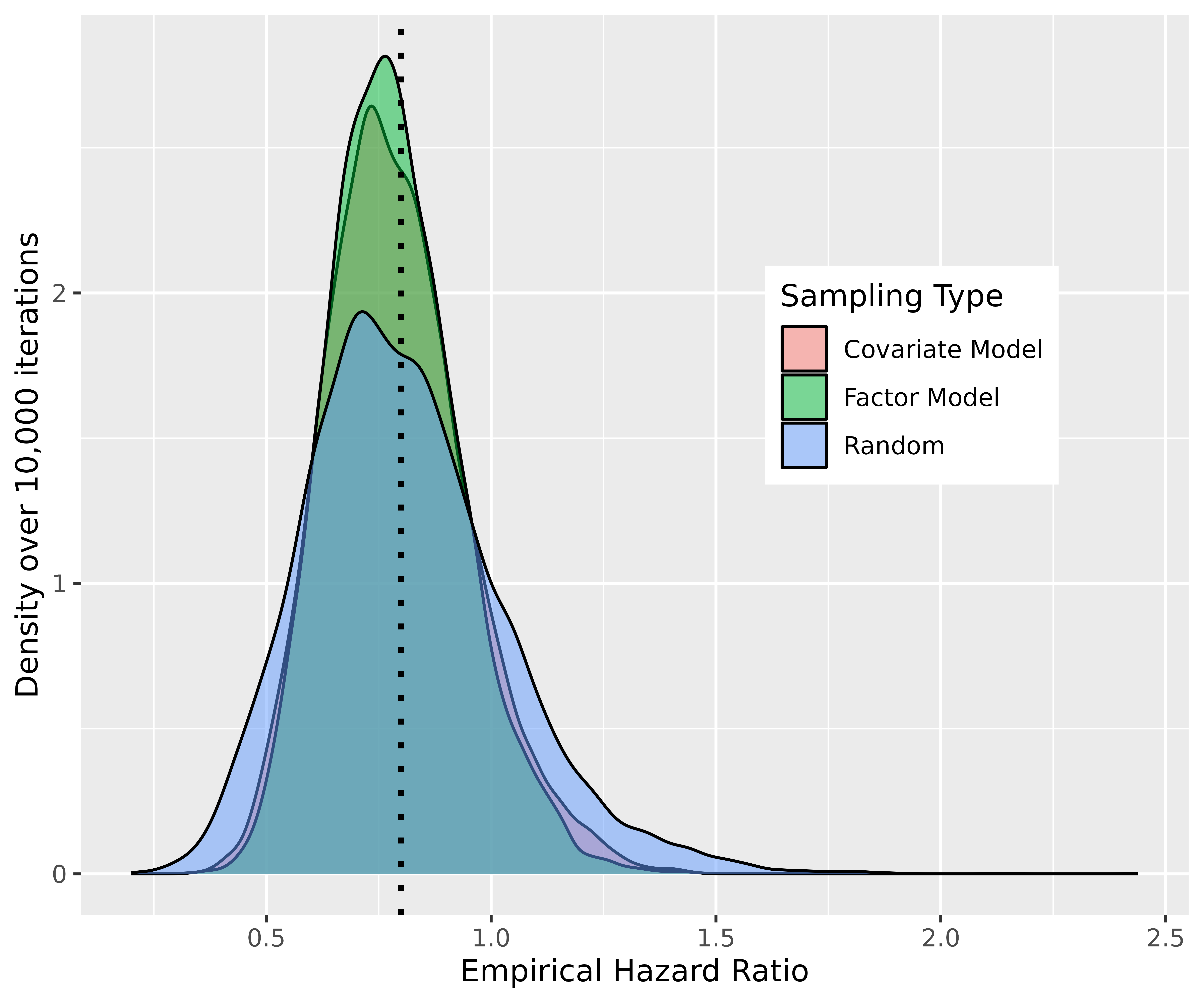}
\caption{Observed Hazard Ratios for a Theoretical Hazard Ratio of 0.8 (Treatment Effect of 0.2)\label{hist}}
\end{figure}

The empirical variability of observed hazard ratios for a theoretical hazard ratio of 0.8 (treatment effect of 0.2) in this simulation is visualized in Figure \ref{hist}. For all three selection methods the distribution of observed effects is somewhat normal and centered around 0.8, demonstrating that the observed treatment effect is an unbiased estimator of the true treatment effect as desired for all three scenarios. We also observe that the variance of the hazard ratios from the random selection method is higher than the other two methods, with the factor model hazard ratios having the lowest variance. These characteristics are also expected since the number of AD events directly correlates with the variance of the effect estimates and  factor model selection should result in the highest number of events while the random selection method should have the lowest.

\begin{figure}[hbt!]
\includegraphics[height=12cm]{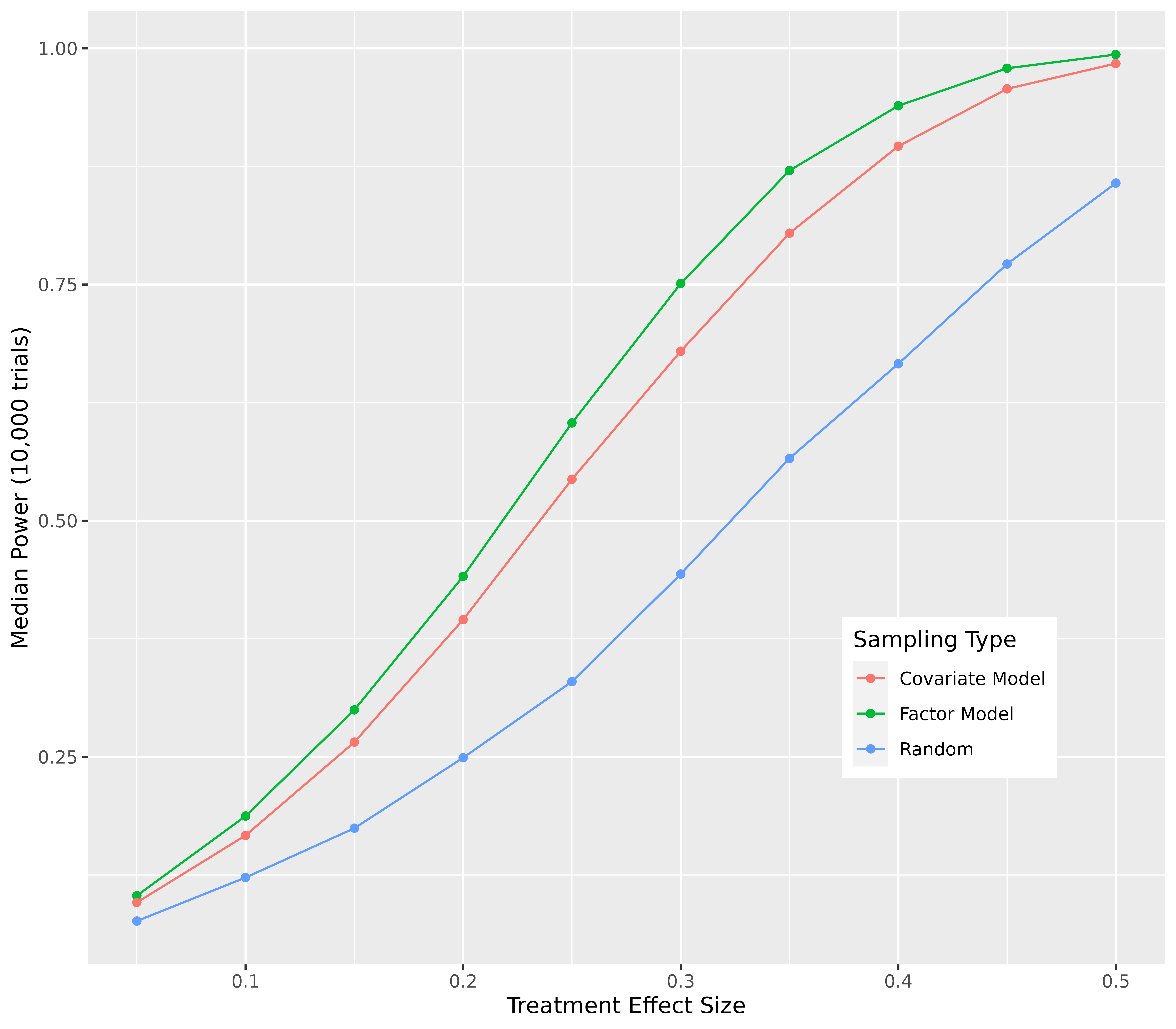}
\caption{Median Power Over a Range of Treatment Effects\label{powerfig}}
 \vspace*{4in}
\end{figure}

A noteworthy consequence of this phenomenon of varying empirical effect estimates is that when examining power relative to the theoretical effect there was substantial variability in the results. Since, for example, the observed hazard ratios for a theoretical hazard ratio of 0.8 may range from about 0.25 to 2.5 (Figure \ref{hist}), the power will appear to vary widely for a theoretical hazard of 0.8. Consequently, Figure \ref{powerfig} displays the median estimated power over the 10,000 trial simulations for each theoretical hazard as it is a stable point estimate that reflects cases where the observed and theoretical hazards are similar.

The results of the power analysis are displayed in Figure \ref{powerfig}. The factor model consistently outperformed the covariates-only model and especially the random selection method. The difference in power was particularly apparent for treatment effects greater than 0.1. For instance, at an effect size of 0.25, the factor model achieved power of 0.60 while the covariate model achieved 0.54 power and the random selection method achieved only 0.33 power. At a treatment effect of 0.3, the factor model achieved about 0.75 power while the covariate model and random selection methods achieved powers of 0.68 and 0.44, respectively.

\begin{figure}[hbt!]
\includegraphics[height=12cm]{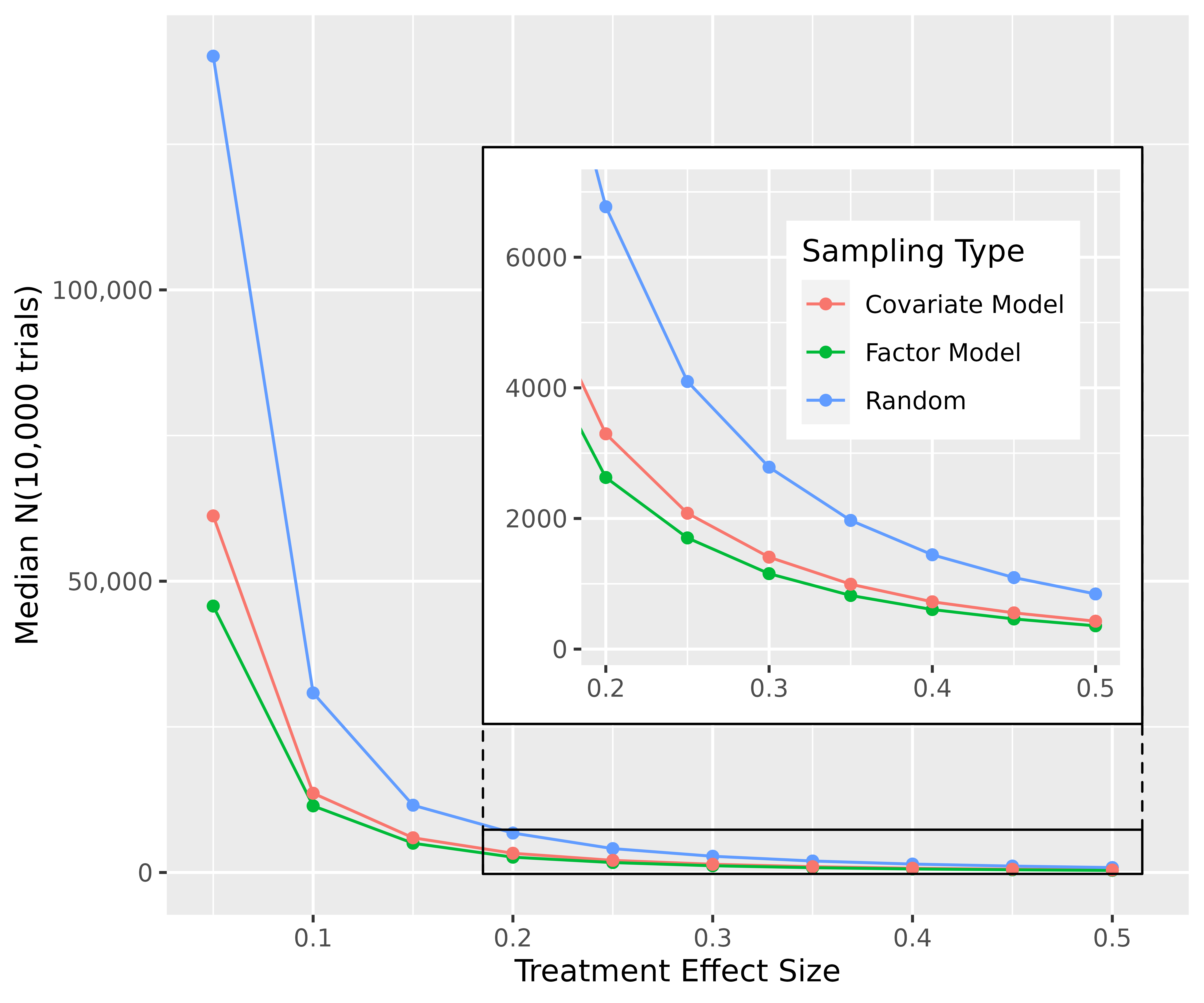}
    \caption{Median N Needed to Recruit Over a Range of Treatment Effects}
    \label{nfig}
     \vspace*{4in}
\end{figure}

The results of the required sample size analysis are displayed in Figure \ref{nfig}. As is expected given the results of the power analysis the factor model outperformed the other two selection options across all treatment effects. For instance, to achieve 80\% power at a treatment effect size of 0.2, the factor model requires about 4,000 less participants than the random selection method and about 500 less participants than the covariate model method. The differences are even more dramatic at lower treatment effects.

\section{Discussion}

We developed a cognitive factor-based model to predict AD conversion from a cognitively normal state within three years. We demonstrated that using this model to facilitate evidence-based subject selection for AD clinical trials would greatly improve statistical power and required sample size over current selection alternatives. It would also allow for enrollment of cognitively normal participants instead of participants with mild cognitive impairment which would greatly increase the enrollment pool as well as allow for testing of a drug that prevents cognitive symptom onset entirely. Further, our model is based only on cognitive test scores and baseline demographics: all data that are more easily and affordably attained than imaging data. Using our predictive model for selection would improve power while also including more patients in trials that do not have access to healthcare, have low socioeconomic status, are racial minorities, and/or are generally a part of a marginalized population. The lack of representation of these participants in recent AD trials severely restricts generalizability and hinders progression towards a widely efficacious AD treatment.

It should be noted that to participate in NACC a patient must have been referred to a memory clinic of some kind. As a result, our simulation very likely overestimates the number of AD events that would occur in a truly random sample of the population (even if age-restricted), and thus overestimates the power. So, the relative improvements in power of the covariate model and our factor model compared to random selection are likely actually larger than what was demonstrated by this analysis.

This analysis has several strengths as well as limitations. Strengths include the relatively large number of AD converters and the novelty of the methods employed. A methodological limitation of this study is the exclusion of incomplete cases, which considerably limits the sample size and applicability of this method to a real-life scenario. We hope to address this concern in future research. Practically speaking, computation time for the factor score extraction alone was quite high: around 36 hours for our sample of around 5,000 participants. Lastly and most importantly, this cohort was overwhelmingly white (84.1\%), which is of particular concern for this research given our hope that these methods be applied specifically to improve racial diversity in AD clinical trials. Further research is required to verify that cognitive factor scores hold the same predictive ability in non-white populations as they do in this cohort.

The use of statistical factor scores derived from a state-space model framework is a novel method that has many uses beyond the scope of this paper. For instance, while this paper performs a complete case analysis, missing test scores occur frequently (see Figure \ref{dataflow}). Factor scores are more robust to imputation methods than individual cognitive test scores and therefore will likely be substantially stronger predictors than individual tests in the presence of missingness. Furthermore, there are a wide variety of cognitive tests with different scoring systems that can still be categorized into the four clinical factors (domains) we use here. Using standardized factor scores would allow for improved harmonization across different tests within the same domain and improve study of cognitive trajectories over time even as the individual cognitive tests change. Finally, to our knowledge this is the only existing method that can derive longitudinal factor scores from repeated measurements on a individual directly, rather than a sequential cross-sectional factor model at each measurement. Furthermore, our model allows for a correlation structure between the factors themselves. We plan to take advantage of these features in future research by jointly modeling longitudinal cognitive factors with time to AD to better understand the longitudinal progression from a cognitively normal state to AD.

In conclusion, we demonstrated the efficacy of a cognitive factor model at improving statistical power and required sample size in a simulated Alzheimer's dementia clinical trial. We hope this method will improve the efficiency of future clinical trials to accelerate the process of finding a much-needed treatment.

\bibliographystyle{plainnat}
\bibliography{bibliography.bib} 

\pagebreak
\section{Tables and Figures}

\begin{table}
\small\sf\centering
\caption{Summary Statistics of Baseline Covariates by AD Status (Test Set Only)}\label{tab1}
\centering
\begin{tabular}[t]{r|rrr}
\toprule
  & \textbf{AD Non-converter} & \textbf{AD Converter} & \textbf{Overall}\\
 & (N=4192) & (N=322) & (N=4514)\\
\midrule
Follow-Up Time (Days) &  &  & \\
Mean (SD) & 527 (462) & 577 (275) & 530 (451)\\
Median [Min, Max] & 412 [0, 1280] & 418 [185, 1280] & 413 [0, 1280]\\
\hline
Sex at Birth &  &  & \\
Male & 1436 (34.3\%) & 120 (37.3\%) & 1556 (34.5\%)\\
Female & 2756 (65.7\%) & 202 (62.7\%) & 2958 (65.5\%)\\
\hline
Years of Education &  &  & \\
Mean (SD) & 15.8 (2.84) & 15.3 (3.12) & 15.8 (2.86)\\
Median [Min, Max] & 16.0 [2.00, 28.0] & 16.0 [2.00, 25.0] & 16.0 [2.00, 28.0]\\
\hline
Age at Baseline &  &  & \\
Mean (SD) & 70.3 (10.1) & 76.6 (8.53) & 70.7 (10.1)\\
Median [Min, Max] & 70.0 [21.0, 99.0] & 78.0 [36.0, 95.0] & 71.0 [21.0, 99.0]\\
\hline
Race &  &  & \\
White & 3508 (83.7\%) & 288 (89.4\%) & 3796 (84.1\%)\\
Black or African-American & 586 (14.0\%) & 28 (8.7\%) & 614 (13.6\%)\\
Am. Indian or Pac. Islander & 14 (0.3\%) & 1 (0.3\%) & 15 (0.3\%)\\
Asian & 84 (2.0\%) & 5 (1.6\%) & 89 (2.0\%)\\
\hline
Number of e4 alleles &  &  & \\
0 e4 alleles & 2953 (70.4\%) & 204 (63.4\%) & 3157 (69.9\%)\\
1 e4 allele & 1142 (27.2\%) & 102 (31.7\%) & 1244 (27.6\%)\\
2 e4 alleles & 97 (2.3\%) & 16 (5.0\%) & 113 (2.5\%)\\
\hline
Hypertension &  &  & \\
Never & 2026 (48.3\%) & 131 (40.7\%) & 2157 (47.8\%)\\
Recent/Active & 2029 (48.4\%) & 179 (55.6\%) & 2208 (48.9\%)\\
Remote/Inactive & 137 (3.3\%) & 12 (3.7\%) & 149 (3.3\%)\\
\hline
Diabetes &  &  & \\
Never/Inactive & 3727 (88.9\%) & 279 (86.6\%) & 4006 (88.7\%)\\
Recent/Active & 465 (11.1\%) & 43 (13.4\%) & 508 (11.3\%)\\
\hline
Smoking Years &  &  & \\
Mean (SD) & 9.81 (14.6) & 11.8 (16.7) & 9.95 (14.8)\\
Median [Min, Max] & 0 [0, 75.0] & 0 [0, 74.0] & 0 [0, 75.0]\\
\hline
BMI &  &  & \\
Non-Obese & 3113 (74.3\%) & 266 (82.6\%) & 3379 (74.9\%)\\
Obese & 1079 (25.7\%) & 56 (17.4\%) & 1135 (25.1\%)\\
\hline
TBI Ever &  &  & \\
No & 3741 (89.2\%) & 284 (88.2\%) & 4025 (89.2\%)\\
Yes & 451 (10.8\%) & 38 (11.8\%) & 489 (10.8\%)\\
\hline
Depression Ever &  &  & \\
No & 3020 (72.0\%) & 217 (67.4\%) & 3237 (71.7\%)\\
Yes & 1172 (28.0\%) & 105 (32.6\%) & 1277 (28.3\%)\\
\bottomrule
\end{tabular}
\end{table}

\begin{table*}[h!]
\small\sf\centering
\caption{Logistic Regression Performance Predicting Three Year AD Risk Using Individual Cognitive Tests Versus Factor Scores (Training Set Only)\label{tenmods}\protect}
\scriptsize
\begin{threeparttable}
\begin{tabular}{c c c c c}
\toprule
Main Predictor & Odds Ratio (95\% CI) & Sensitivity & Specificity\\
\midrule
Immediate Recall\tnote{*} &  0.67 (0.58, 0.77) & 0.70 & 0.66 \\
Delayed Recall\tnote{*} &0.60 (0.52, 0.69)  & 0.70 & 0.68 \\
\hdashline
Memory Factor\tnote{*} & 0.55 (0.47, 0.64) & 0.71 & 0.67 \\
\hline
Digit Span Forward\tnote{*} & 0.82 (0.71, 0.94)  & 0.69 & 0.65\\
Digit Span Backward & 0.88 (0.76, 1.01) & 0.69 & 0.67\\
\hdashline
Working Memory Factor\tnote{*} &  0.83 (0.72, 0.95) & 0.69 & 0.67\\
\hline
Animal List Generation\tnote{*} & 0.67 (0.57, 0.78) & 0.70 & 0.67\\
Vegetable List Generation\tnote{*} & 0.63 (0.53, 0.73) & 0.69 & 0.67\\
Boston Naming Test\tnote{*} & 0.72 (0.64, 0.83) & 0.70 & 0.70\\
\hdashline
Language Factor\tnote{*}& 0.59 (0.51, 0.68) & 0.71 & 0.67\\
\hline
Trail Making Test Part A\tnote{*} & 0.79 (0.70, 0.89) & 0.71 & 0.66 \\
Trail Making Test Part B\tnote{*} & 0.83 (0.73, 0.94) & 0.70 &0.67\\
Digit Symbol\tnote{*} & 0.65 (0.56, 0.76) & 0.69 & 0.68\\
\hdashline
Psychomotor Speed Factor\tnote{*} & 0.68 (0.60, 0.78) & 0.71 & 0.68\\
\bottomrule
\end{tabular}
\begin{tablenotes}
    \item[] Each cognitive test loads onto the factor at the end of its respective table section. All models are adjusted for sex, education, age at baseline, race, APOE4, hypertension, diabetes, smoking years, binary BMI, TBI history, and depression.
    \item[*] Statistically significant at the 0.05 level.
\end{tablenotes}
\end{threeparttable}
\end{table*}

\end{document}